\documentclass{article}
\usepackage{arxiv}
\usepackage[T1]{fontenc}    
\usepackage{amsmath}
\usepackage{amssymb}
\usepackage{booktabs}       
\usepackage{amsfonts}       
\usepackage{nicefrac}       
\usepackage{microtype}      
\usepackage{lipsum}		
\usepackage{graphicx}
\usepackage{doi}
\usepackage{subfiles}
\usepackage[linesnumbered,lined,boxed,commentsnumbered]{algorithm2e}
\usepackage{caption}
\usepackage{subcaption}
\usepackage{cancel}

\DeclareMathOperator*{\argmax}{argmax}
\DeclareMathOperator*{\argmin}{argmin}

\title{Optimal Settings \\ for Cryptocurrency Trading Pairs}

\author{ 
	{Di Zhang\textsuperscript{a}, Youzhou Zhou\textsuperscript{b}} \\
	{School of AI and Advanced Computing\textsuperscript{a}, School of Mathematics and Physics\textsuperscript{b}}\\
	Xi'an Jiaotong-Liverpool University\\
	Suzhou, 215123, China PR \\
	\texttt{\{di.zhang, youzhou.zhou\}@xjtlu.edu.cn}
}

\date{}

\begin{document}
	\maketitle
	\begin{abstract}
		
		
		The goal of cryptocurrencies is decentralization, but it is impractical to set up a trading market between every two currencies. To solve this optimization problem, we use a two-stage process: 1) Fill in missing values based on a regularized, truncated eigenvalue decomposition, where the regularization term is used to control what extent missing values should be limited to zero. 2) Search for the optimal trading pairs, based on a branch and bound process, with heuristic search and pruning strategies. 
		
		The experimental results show that: 1) If the number of denominated coins is not limited, we will get a more decentralized trading pair settings, which advocates the establishment of trading pairs directly between large currency pairs. 2) There is a certain room for optimization in all exchanges. The setting of inappropriate trading pairs is mainly caused by subjectively setting small coins to quote, or failing to track emerging big coins in time. 3) Too few trading pairs will lead to low coverage; too many trading pairs will need to be adjusted with markets frequently. Exchanges should consider striking an appropriate balance between them.

	\end{abstract}

	\section{Introduction}
	
	In the crypto market, people can freely issue and trade virtual coins\cite{Nakamoto2008BitcoinAP}\cite{Wood2014ETHEREUMAS}. These cryptocurrencies are growing rapidly in number, differ widely in market cap, and change frequently. Traditionally, people are accustomed to denominated in a fiat currency (such as the US dollar) and exchange assets with it to achieve asset conversion. This is the so-called "universal equivalence". However, in the age of cryptocurrencies, the absence of a unified, constant, mandatory pricing benchmark has made the trading process either simpler, or even more cumbersome. For example, suppose a trader wants to exchange asset A for B. In the traditional stock market, this requires 2 exchanges as the fiat is needed for transit. In the crypto market, due to the different settings of the trading pair, this number may be 1, i.e., direct exchange; or more than 2 times, i.e., multi-jump exchange.
	
	Obviously, the setting of trading pairs has a significant impact on the efficiency of trading. If the number of trading pairs is insufficient, or if the settings are inappropriate, it will increase the user's transaction costs (slippage + transaction fee). Moreover, the price of transit currencies does not always remain stable. Even so-called stablecoins can be affected by unexpected events that lead to  disanchoring, such as the LUNA/UST crash\footnote{https://insidebitcoins.com/news/explaining-the-luna-crash}. This may result in unexpected losses for users during the transit process. On the other hand, the number of trading pairs should not be excessive. First, too many trading pairs can diversify liquidity, leading to wider market spreads and higher transaction costs. Second, if the number of trading pairs grows at the rate of $\mathcal{O}(N^2)$ given $N$ as the number of coins, then maintenance can be extremely expensive and not feasible in practice.
	
	This becomes an optimization problem, i.e., to cover maximal volume when the intentional volume matrix is $\mathbf{K}^*$ and the trade pairs can be set upto $M$. This problem is similar to the Origin-Destination Matrix (ODM) estimation problem in transportation planning\cite{ros2022practical}. In simplification, it can be considered as a missing value filling problem. The difficulty is that we have a very high proportion of missing values here, and applying the existing model directly will most likely lead to overfitting of a small number of observations and overprediction of missing values. To prevent this, we add a regularization term to the usual eigenvalue decomposition\cite{Cohen2021ATO} to confine missing values around 0. After obtaining the estimate of $\mathbf{K}^*$, due to the need to ensure that all coins are reachable, we construct a branch and bound method, which searches for the optimal setting by a heuristic search, with the corresponding pruning and early termination strategies.	
	
	The experimental results show that based on the decomposition of eigenvalues, we can obtain a result with good interpretability, which is helpful to understand the mechanism of volume formation. Moreover, by running the optimization algorithm, it is possible to scan out some unreasonable trading pair settings in the exchange. These pairs do show a scarcity of trading volumes in practice. In addition, by running the crypto market sequentially in the most recent year, we found that the number of trading pairs should be maintained within a reasonable range, which can both ensure the efficiency of market trading and maintain sufficient stability in the volatile change in the volume ranking. 
	



\section{Related Work}

There are two areas that are closely related to this paper: ODM and triangular arbitrage.

The estimation of ODM is a fundamental problem in transportation planning\cite{egu2020comparable}. Accurate ODMs provide valuable information about the movement of people and goods between various locations which is used to understand travel patterns, forecast future demand, and design efficient transportation systems. The direct approach of collecting data on the movement of people or goods, such as household surveys, traffic counts, or GPS data, is one of the commonly used methods to estimate the ODM\cite{krishnakumari2020data}. The indirect approach of using external information, such as land use data, transportation network data, and socio-economic data, is often used when direct data is not available or is too expensive to collect\cite{fekih2021data}. 

Triangular arbitrage is a popular trading strategy in the cryptocurrency market due to the high volatility and liquidity of this market. The strategy involves purchasing one cryptocurrency using another cryptocurrency, then selling that cryptocurrency to purchase a third cryptocurrency. The process is then repeated in reverse to take advantage of the price differences between the three currencies\cite{fang2022cryptocurrency}. Other strategies in the crypto market that take advantage of the correlation between currencies include pair trading\cite{ramos2017introducing}\cite{ramos2020some}\cite{fil2020pairs}, derivatives arbitrage, etc\cite{fang2021ascertaining}.

In general, this article attempts to consider how to improve transaction efficiency from the perspective of exchange operators, which is different from all the above work. To some extent, exchange operators and arbitrageurs are thinking in opposite directions. For example, when there is a better path than direct transactions between multiple currencies, the exchange may suggest users to perform multi-step conversions, which cancels arbitrage opportunities in advance.


 





	\section{Problem Definition}\label{sec:def}
	
	Assuming that the given number of coins is $N$, and $M$ trading pairs are defined between them, this constitutes a symmetric undirected graph $\mathbf{G}$, where the weight is $0$ or $1$. In this case, we observe a volume matrix $\mathbf{V}$, which is also a symmetric matrix, where $V_{i,j}>0$. Note that $V_{i,j}=0$, if $G_{i,j}=0$. We hope to reconstruct $\mathbf{G}$ as $\mathbf{G}^*$, so that the sum of volume $\mathbf{V}^*=\mathbf{K}^*\circ \mathbf{G}^*$ is maximized. Note that $\mathbf{G}^*$ must be connected. This problem can be formally written as
	\begin{equation}\label{eqn:set}
		\begin{aligned}
	\argmax_{\mathbf{G}^*}& \; \mathbf{K}^*\circ \mathbf{G}^* \\
	\text{s.t. }& G^*_{i,j}=G^*_{j,i}, \; G^*_{i,j}\in\{0,1\}, \; \text{for } i,j\in \{1,\ldots,N\} \\
	&\; \mathbf{G}^* \text{ is connected}.
\end{aligned}
	\end{equation}
Among them, $\mathbf{K^*}$ is the intentional volume of trading to be estimated. As a symmetric matrix, we assume that its rank is only $2$, and the eigenvalues are one positive and one negative (refer to the analysis of the data in Sec. \ref{sec:data}), then the decomposition of the following form can be obtained: $\mathbf{K }^*\triangleq\mathbf{w}_1\mathbf{w}_1^T-\mathbf{w}_2\mathbf{w}_2^T$, where $\mathbf{w}_1$ and $\mathbf{ w}_2$ is orthogonal.
		\begin{equation}\label{eqn:odm}
\begin{aligned}
	\argmin_{\mathbf{w}_1,\mathbf{w}_2} & \left \| \mathbf{G}\circ \left(\mathbf{K}^* - \mathbf{K} \right) \right \|_F^2+\lambda \left \| \left(\mathbf{1}-\mathbf{G}\right)\circ \mathbf{K}^* \right \|_F^2, \\
	\text{s.t. } &\mathbf{w}_1^T \; \mathbf{w}_2=0\\
	& \mathbf{w}_{i,1}\geq 0,\;\text{for } i\in \{1,\ldots,N\} \\
	& K_{i,j}\geq 0,\;\text{for } i,j\in \{1,\ldots,N\}
\end{aligned}
\end{equation}
where $\mathbf{K}$ is the realized volume; $\mathbf{1}$ is the full $1$ matrix of the same size as $\mathbf{G}$; $\lambda$ is the regularization coefficient. We expect $\mathbf{w}_1$ to represent the "mass" of the node (see ODM's gravity model\cite{Wang2020ResearchOR}, which analogizes the amount of flow to the magnitude of gravity). In order to make this meaning clearer, we further require all elements of $\mathbf{w}_1$ are required to be not less than 0. However, we do not limit the signs of $\mathbf{w}_2$. If the two elements in $\mathbf{w}_2$ have the same sign, it indicates that there is an "repulsion" between them, whereas the different sign represents an "suction".

Note that the following effects are not considered here: 1) The detour effect: when there is no trading pair between the two coins, other trading pairs must be detoured, resulting in an increase in their traffic. 2) The aggregation effect: Since the market spread for large trading pairs is narrower, the transaction cost of detouring is even lower than that of direct trading between small trading pairs. We assessed these effects can be meaningful for the further optimization, but not decisive at current stage.

	\section{Solution}
	\subsection{Estimate of $\mathbf{K}^*$}\label{sec:algok}
	We first solve target \ref{eqn:odm} using the interior point method\cite{pastor2005solving}, which is easy to understand and has fast convergence. For clarity, denote the optimization objective in \ref{eqn:odm} as $f(\mathbf{W})$, where $\mathbf{W}\triangleq \left[\mathbf{w}_1\;\mathbf {w}_2\right]$, and record the equality constraints as $g(\mathbf{W})$, and the two inequalities as $h_i^{(1)}(\mathbf{W}), h_{i,j}^{(2)}(\mathbf{W})$. It can be approximated by the following problem:
	\begin{equation}
		\begin{aligned}
					\min_{\mathbf{W},\mathbf{s} } & f(\mathbf{W})-\lambda \sum_i \ln(s_i),\\
			\text{s.t. } & s_i\geq 0,\;g(\mathbf{W})=0,\;h_i(\mathbf{W})+s_i=0.
		\end{aligned}
	\end{equation}
Its Lagrangian function is
	\begin{equation}\label{eq:la}
	\begin{aligned}
		\mathcal{L}(\mathbf{W},\mathbf{s},\alpha,\beta)=f(\mathbf{W})-\lambda \sum_i \ln(s_i)-\alpha g(\mathbf{W})-\boldsymbol{\beta}^T \left(\mathbf{h}(\mathbf{W})-\mathbf{s}\right),
	\end{aligned}
\end{equation}
where $\alpha,\boldsymbol{\beta}$ are Lagrange multipliers. Next, we solve the Eq. \ref{eq:la} using Newton's iteration method. We pay particular attention to $f_E(\mathbf{G},\mathbf{K})\triangleq\mathbf{G}\circ \left(\mathbf{K}^* - \mathbf{K} \right)^{\circ 2}$ in the first term of \ref{eqn:odm}, its partial derivative at position ${i,j}$ to $W_{k,l}$ is
	\begin{equation}
		\begin{aligned}	
			\tilde{J}_{i,j,k,l}(G_{i,j},K_{i,j})=2&*G_{i,j}*\overbrace{\left( W_{1,i}W_{1,j}-W_{2,i}W_{2,j}-K_{i,j} \right)}^{\textcircled{A}} \\
		&* \underbrace{ \left[ \mathbb{I}(k=1,l=i)W_{1,j} + \mathbb{I}(k=1,l=j)W_{1,i} - \mathbb{I}(k=2,l=i)W_{2,j} - \mathbb{I}(k=2,l=j)W_{2,i} \right] }_{\textcircled{B}},
	\end{aligned}
\end{equation}
The second-order partial derivative of $W_{m,n}$ at position ${i,j}$ to $W_{k,l}$, $W_{m,n}$ is
	\begin{equation}
	\begin{aligned}
		\tilde{H}_{i,j,k,l,m,n}(G_{i,j},K_{i,j})=&2*G_{i,j}*\textcircled{B}*[ \mathbb{I}(m=1,n=i)W_{1,j} + \mathbb{I}(m=1,n=j)W_{1,i} \\
		&-\mathbb{I}(m=2,n=i)W_{2,j} - \mathbb{I}(m=2,n=j)W_{2,i} ]\\
		+&2*G_{i,j}*\textcircled{A}*[ \mathbb{I}(k=1,l=i,m=1,n=j) + \mathbb{I}(k=1,l=j,m=1,n=i) \\
		&- \mathbb{I}(k=2,l=i,m=2,n=j) - \mathbb{I}(k=2,l=j,m=2,n=i) ].
	\end{aligned}
\end{equation}
After summing up, we get the Jacobian and Hessian of $f_E$ as
\begin{equation}
	\tilde{J}(\mathbf{G},\mathbf{K})_{k,l}=\sum_{i=1}^N \sum_{j=1}^N \tilde{J}_{i,j,k,l},
\end{equation}
\begin{equation}
 \tilde{H}(\mathbf{G},\mathbf{K})_{k,l,m,n}=\sum_{i=1}^N \sum_{j=1}^N \tilde{H}_{i,j,k,l,m,n}.
\end{equation}
Therefore, the Jacobian and Hessian of $f$ as a whole are
\begin{equation}\label{eq:j}
	J_{k,l}=\tilde{J}_{k,l}(\mathbf{G},\mathbf{K})+\lambda*\tilde{J}_{k,l}(\mathbf{1}-\mathbf{G},\mathbf{0}),
\end{equation}
\begin{equation}\label{eq:h}
	H_{k,l,m,n}=\tilde{H}_{k,l,m,n}(\mathbf{G},\mathbf{K})+\lambda*\tilde{H}_{k,l,m,n}(\mathbf{1}-\mathbf{G},\mathbf{0}).
\end{equation}
Meanwhile, the partial derivatives of $g$, $h_i^{(1)}$ and $h_{i,j}^{(2)}$ to $W_{k,l}$ are:
\begin{equation}\label{eq:jg}
J^{g}_{k,l}=\mathbb{I}(l=1)W_{k,2}+\mathbb{I}(l=2)W_{k,1},
\end{equation}
\begin{equation}\label{eq:jh1}
	J^{h^{(1)}}_{l}=\mathbb{I}(l=1),
\end{equation}
\begin{equation}\label{eq:jh2}
	J^{h^{(2)}}_{k,l}=\textcircled{B}.
\end{equation}
The second order partial derivatives of the above three quantities are all 0. Then, we can list the following equation to calculate the gradients used in Newton's method\cite{byrd2000trust}:
\begin{equation}\label{eq:new}
\begin{bmatrix}
	\hat{\mathbf{H}} & \mathbf{0} &  (\hat{\mathbf{J}}^g)^T & (\hat{\mathbf{J}}^h)^T \\ 
	\mathbf{0} &  \mathbf{S} \mathbf{A}   & \mathbf{0} & \mathbf{0}\\ 
	\hat{\mathbf{J}}^g & \mathbf{0} & 0 & \mathbf{0}\\ 
	\hat{\mathbf{J}}^h &  \mathbf{S} & \mathbf{0} & \mathbf{0}
\end{bmatrix}\begin{bmatrix}
	\Delta \hat{\mathbf{w}}\\ 
	\mathbf{S}^{-1}\Delta \mathbf{s}\\ 
	\Delta \alpha\\ 
	\Delta \boldsymbol{\beta}
\end{bmatrix}=\begin{bmatrix}
	\hat{\mathbf{J}}+ (\hat{\mathbf{J}}^g)^T\alpha+(\hat{\mathbf{J}}^h)^T\boldsymbol{\beta} \\ 
	\mathbf{S}\boldsymbol{\beta}-\lambda\mathbf{1}\\ 
	g\\ 
	\mathbf{h}+\mathbf{s}
\end{bmatrix}
\end{equation}
To express more succinctly, we rearrange $\mathbf{W}$ into a one-dimensional vector $\hat{\mathbf{w}}=vec_{2N\times 1}(\mathbf{W})$, and rearrange the corresponding formula \ref{eq:j},\ref{eq:h},\ref{eq:jg},\ref{eq:jh1} and \ref{eq:jh2}, i.e. $\hat{\mathbf{J}} =\text{vec}_{2N\times 1}(\mathbf{J})$,$\hat{\mathbf{H}}=\text{vec}_{2N\times 2N}(\mathbf{H })$, $\hat{\mathbf{J}}^g=\text{vec}_{1\times 2N}(\mathbf{J}^g)$, $\hat{\mathbf{J}} ^h=\text{vec}_{ (N^2+N)\times 2N}([\mathbf{J}^{h^{(1)}}\;\mathbf{J}^{h^{ (2)}}])$; in addition, $\mathbf{S} = \text{diag}(\mathbf{s}), \mathbf{A} = \text{diag}(\alpha)$. The Eq. \ref{eq:new} can be quickly solved by Cholesky Decomposition, and the resulting $\Delta \hat{\mathbf{w}}$, $\Delta \mathbf{s}$ are used for Newton iteration. In multiple iterations, we continuously reduce $\lambda$ and obtain the required $\mathbf{K}^*$ after $\lambda$ is small enough.

\subsection{Search for $\mathbf{G}^*$}

Without considering connectivity constraints, the formula \ref{eqn:set} can be quickly solved using the greedy method. But to guarantee connectivity, we need to use a more sophisticated approach. Here we use the branch and bound method \cite{tomazella2020comprehensive}, which divides the feasible space as $\mathcal{S}=\{\mathbf{G}\}$ into disjoint subsections with edges and no edges at each position, and compute the upper and lower bounds of the solutions in each subset. It always divides the subspace that is most promising (i.e., the upper bound is largest) to produce a better solution, and terminates the search in the subspace where it is no longer possible to produce a better solution (i.e., the lower bound is higher than the current optimal). We note that if the number of edges in $\mathbf{G}$ must not exceed $M$, then the number of edges in the optimal solution must be $M$; otherwise, we can always add edges to get a better solution. Therefore, when dividing the subspace, we can prematurely terminate those subsets whose remaining available edges are less than $M$, or whose remaining empty edges have exceeded $\frac{N(N-1)}{2}-M$. This practice can significantly speed up the search process. The main process of the algorithm is summarized in \ref{alg:opt}.

\begin{algorithm}[!htbp]
	
	\SetAlgoLined

%
	q $\gets$ init\_q('edges','lb','ub','visited'); \tcp{Initialize the queue, including edge coverage, upper and lower bounds, and whether it has been visited. Add all NaNs as the root node}
	\While{true}{
		\tcp{Put the unvisited ones with the largest upper bound to the top}
		q $\gets$ sort(q)\;
		\eIf{$q_1$.visited}{
			break;
		}{
\tcp{If the upper bound is not as good as the known best lower bound, then prune}
		\If{~cutable($q_1$,q)}{ 
				$q_N$ $\gets$ expand($q_1$); \tcp{Take an empty position, try with or without edges, generate two nodes}
				
				\tcp{Fill empty edges in advance}
				\If{$\sum q_N$.edges==$M$}{
					$q_N$.edges(isnan($q_N$.edges))=0; \tcp{The number of edges is used up, and the rest are all set to no edges}
				} 
			\If{$\sum \neg q_N$.edges==$M$}{
				$q_N$.edges(isnan($q_N$.edges))=1; \tcp{There can be no more edges, the rest are all set to filled}
			}
				\BlankLine
				\tcp{Estimate upper bound}
				$q_N\text{.ub} \gets \sum \mathbf{G}_{q_N.edges}\circ\mathbf{K}^*+\sum \text{top}_{N*(N-1)/2-M} \left(\mathbf{G}_{isnan(q_N.edges)}\right)\circ\mathbf{K}^*$;\tcp{Already covered, plus the maximum possible coverage in the future}
				
				\BlankLine
				\tcp{Check lower bound}
				\eIf {$\sum$q.edges==$M$ \& isconnected(q.edges)} {
					$q_N$.lb $\gets$ $q_N$.ub;
				} {
				$q_N$.lb $\gets$ 0;
				}
				$q_N$.visited $\gets$ false; \tcp{not yet visited}
				
				q $\gets$ add\_q(q, $q_N$); \tcp{join the queue}
			}
		
		$q_1$.visited $\gets$ true \tcp{already visited}
	}
}
\caption{Basic process of searching $\mathbf{G}^*$ based on branch and bound method.}
\end{algorithm}\label{alg:opt}

	\section{Experiments}
	\subsection{Data}\label{sec:data}
\begin{table}[htbp]
	\centering
		\caption{Dataset Overview.}
\begin{tabular}{@{}lllllll@{}}
	\toprule
	ID & Period    & Exchange & \#Coins & \#Pairs & Volume(\$)     & Top20(\%) \\ \midrule
	1  & 21.7-22.7 & Binance  & 391    & 1474   & 2.0798e+11 & 78.53       \\
	2  & 21.7-22.7 & FTX      & 416    & 543    & 2.7252e+10 & 94.70       \\
	3  & 21.7-22.7 & KuCoin   & 556    & 746    & 1.2842e+10 & 80.14       \\ \bottomrule
\end{tabular}
\label{tab:data}
\end{table}

We collected the transaction volume data of the three most representative exchanges (Binance\footnote{https://www.binance.com/}, FTX\footnote{https://www.ftx.com/, bankrupted shortly after this article was written}, KuCoin\footnote{ https://www.kucoin.com/}) in the last year, as shown in the Tab. \ref{tab:data}. It can be seen that Binance has the largest number of trading pairs, which is about $3\sim 4$ times the number of coins; while the other two exchanges are even more parsimonious, only about 30\% more than the number of coins. If we only extract the top 20 major currency-related trading pairs, they can account for 78\%-94\% of the total trading volume. Therefore, the distribution of transaction volume is quite uneven, and the optimization of major currencies is the key to solving this problem. For ease of visualization, we will use only the top20 coins in Sec. \ref{sec:estk} and \ref{sec:optset}; in Sec. \ref{sec:timevar}, we will use all trading pairs.

\begin{figure}[htbp]
	\centering	     
			\includegraphics[width=0.8\linewidth]{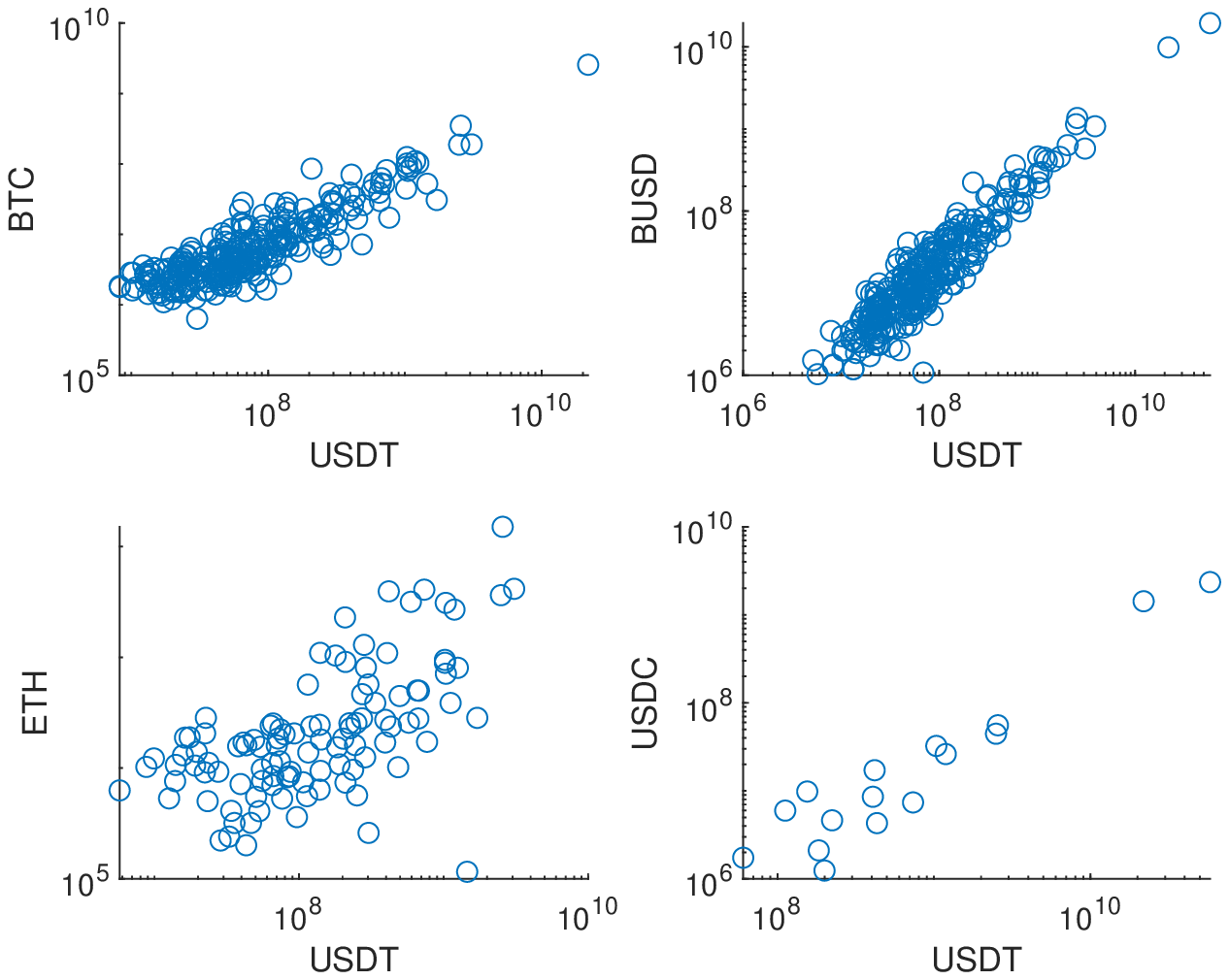}
			\caption{The trading volume corresponding to different quotes and the same base in Binance.}
			\label{fig:scq}
\end{figure}
In addition, in order to illustrate that the low-rank assumption of the traffic matrix in \ref{sec:def} is basically reasonable, the transaction volume corresponding to the same base under different quotes is given in Fig. \ref{fig:scq}. They show a significant linear relationship. This shows that only using $\mathbf{w}_1\mathbf{w}_1^T$ in \ref{eqn:odm}, that is, the gravity model, can roughly reflect the basic situation of flows. After adding $-\mathbf{w}_2\mathbf{w}_2^T$, it can be further corrected. As for higher-order relationships, there seems to be no need to explore further.

	\subsection{Estimation of $\mathbf{K}^*$}\label{sec:estk}
	Using the method in \ref{sec:algok}, we estimate the $\mathbf{w}_1$ and $\mathbf{w}_2$ corresponding to the dataset in the Tab. \ref{tab:data}, as shown in the Fig. \ref {fig:wb}, \ref{fig:wb} and \ref{fig:wb}. Empirically, we can fix $\lambda=0.5$ in \ref{eqn:odm}. Only if this value is not too extreme, it has a limited impact on the final result. 
	
	It can be seen from the Fig. \ref{fig:wb},\ref{fig:wb} and \ref{fig:wb} that the "mass" of the coins displayed by $\mathbf{w}_1$ is indeed related to daily experience Yes, the larger the value, the more often it is used as a transaction transfer. Moreover, the situation of the three exchanges is basically similar, and the differences may be caused by exchange policies (USD is higher than USDT on FTX), or trading habits (ETH is higher than BTC on FTX and Kucoin). The "repulsion" of the coin displayed by $\mathbf{w}_2$ is also in line with the actual observation. Between different large currency transactions, such as BTC and ETH, the number of direct transactions is always low; and the exchange demand between stablecoins is often less. This effect cannot be ignored, but the cause is not clear and needs further study.
	
It can be seen from the Fig. \ref{fig:wb},\ref{fig:wb} and \ref{fig:wb} that the "mass" of the coins displayed by $\mathbf{w}_1$ is indeed meets the daily experience. The higher the value, the more often it is used as a transfer. Moreover, the "mass" of coins on the three exchanges is basically similar, and the differences may be caused by exchange policies (USD is higher than USDT on FTX), or trading habits (ETH is higher than BTC on FTX and Kucoin). The "repulsion" of the coin displayed by $\mathbf{w}_2$ is also in line with the actual observation. Between different big coins, such as BTC and ETH, the number of direct transactions is always lower than expected; and the coversion between stable coins is also often less. This effect cannot be ignored, although the cause is unclear and requires further study.

	\begin{figure}[htbp]
		\centering	     
		\includegraphics[width=0.99\linewidth]{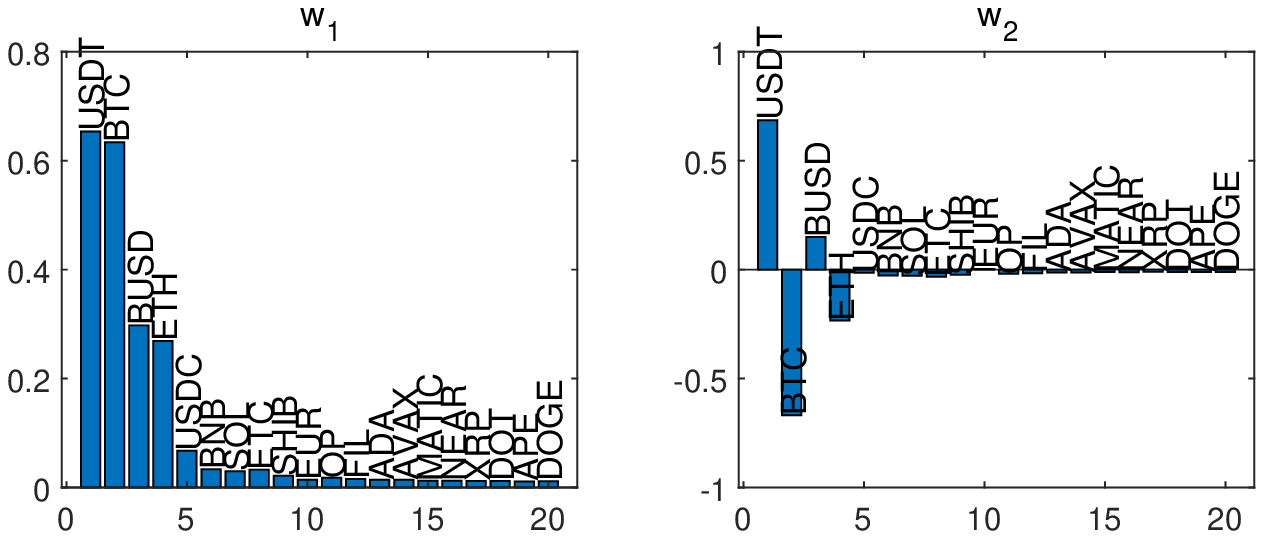}
		\caption{$\mathbf{W}$ of Binance.}
	\end{figure}\label{fig:wb}

	\begin{figure}[htbp]
	\centering	     
	\includegraphics[width=0.99\linewidth]{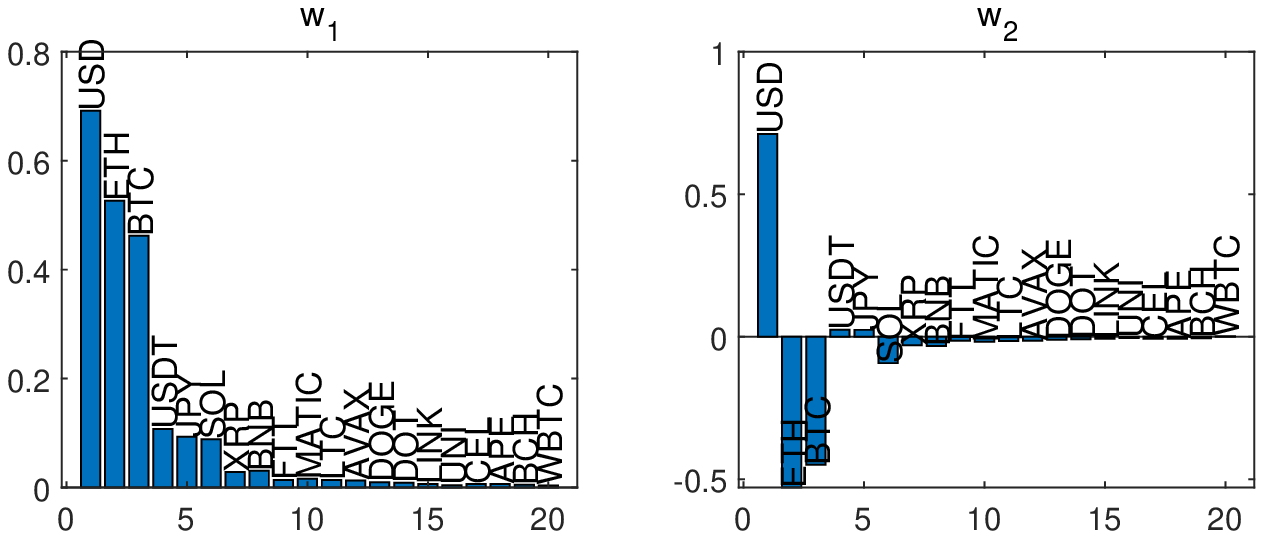}
	\caption{$\mathbf{W}$ of FTX.}
\end{figure}\label{fig:wf}

	\begin{figure}[htbp]
	\centering	     
	\includegraphics[width=0.99\linewidth]{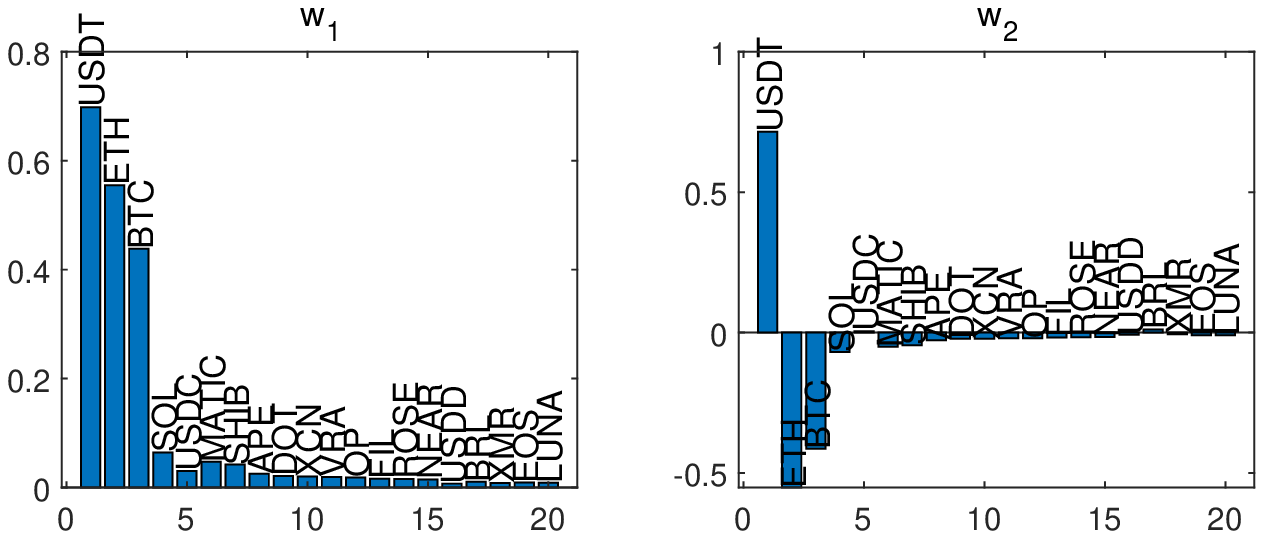}
	\caption{$\mathbf{W}$ of Kucoin.}
\end{figure}\label{fig:wk}

\subsection{Optimal Trading Pairs}\label{sec:optset}

	\begin{figure}[htbp]
	\centering	     
	\includegraphics[width=0.99\linewidth]{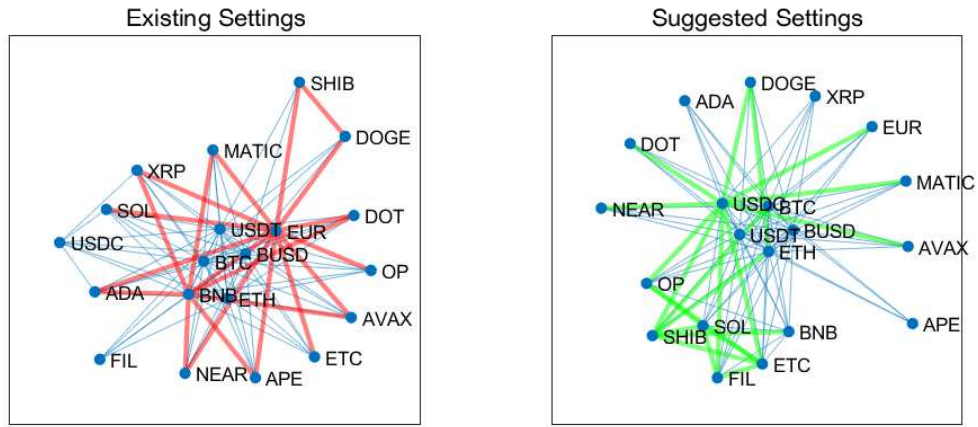}
	\caption{Binance's trading pairs before and after optimization. Red means removed, green means newly added.}
\end{figure}\label{fig:s1}

	\begin{figure}[htbp]
	\centering	     
	\includegraphics[width=0.99\linewidth]{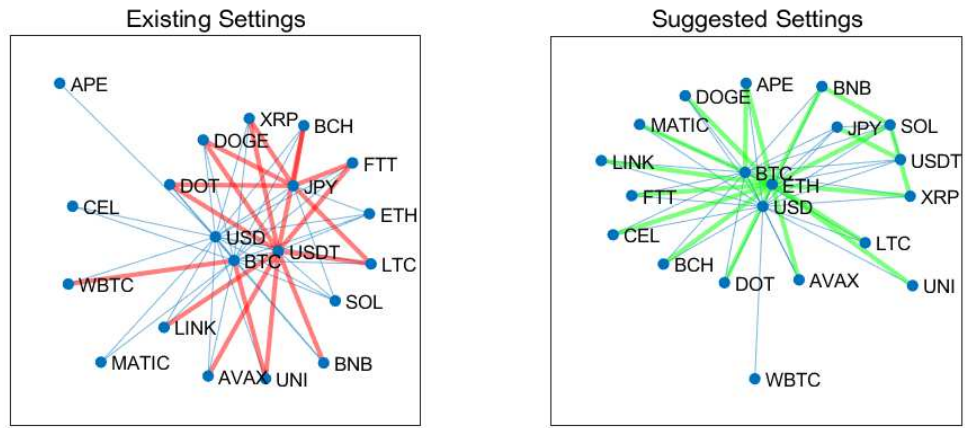}
	\caption{FTX's trading pairs before and after optimization.}
\end{figure}\label{fig:s2}

	\begin{figure}[htbp]
	\centering	     
	\includegraphics[width=0.99\linewidth]{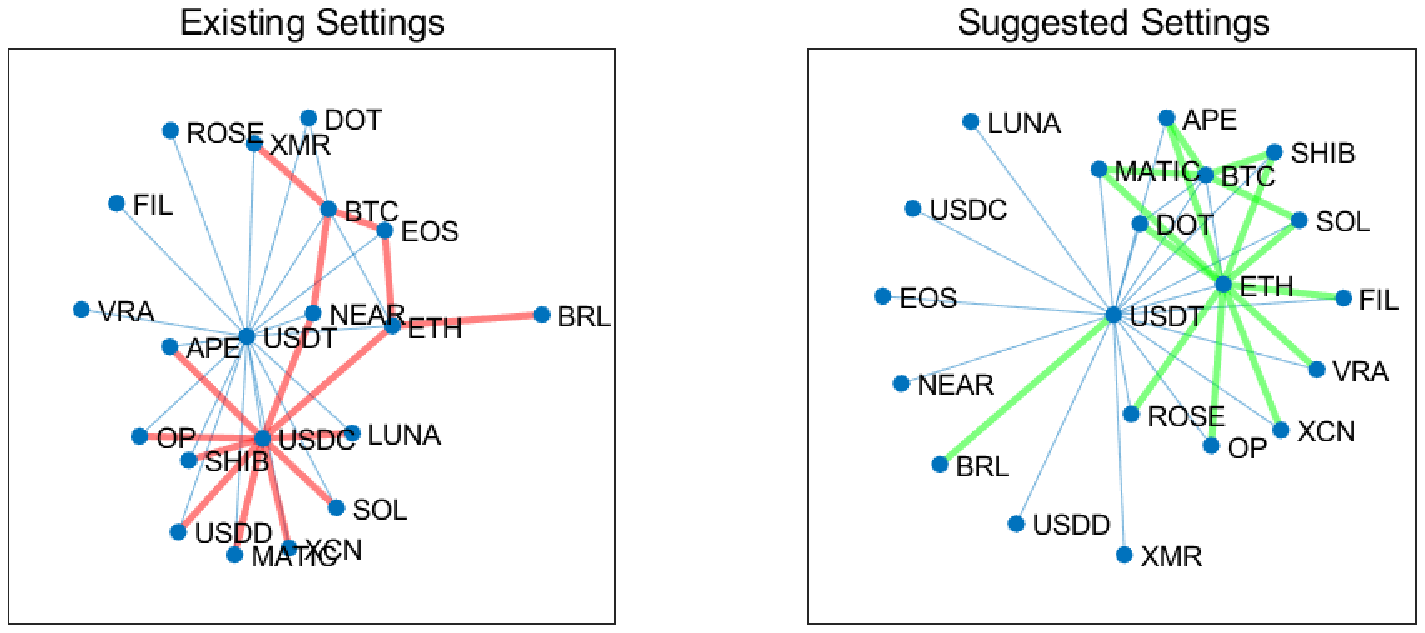}
	\caption{Kucoin's trading pairs before and after optimization.}
\end{figure}\label{fig:s3}

After running Algo. \ref{alg:opt} on the most recent volume data in Tab. \ref{tab:data}, we get data in Fig. \ref{fig:s1}, \ref{fig:s2} and \ref{fig:s3}. In the experiment, we keep the number of trading pairs $M$ constant. We observed that some quotes artificially designated by exchanges, such as Binance's BNB and EUR, were removed by the algorithm and replaced by more mainstream currencies such as ETH. A similar situation also occurs in quotes such as USDT and JPY of FTX, USDC of KUcoin, etc. It is true that exchanges have a subjective factor in promoting these coins, but the transactional efficiency of the market may be hindered. In addition, there are some newly added trading pairs on emerging coins, such as Binance's SOL, etc. In view of the drastic changes in the trading volume in the market, there may be untimely adjustments under current manual control.

\subsection{Changes of Setting over Time}\label{sec:timevar}


	\begin{figure}[htbp]
	\centering	     
	\includegraphics[width=0.9\linewidth]{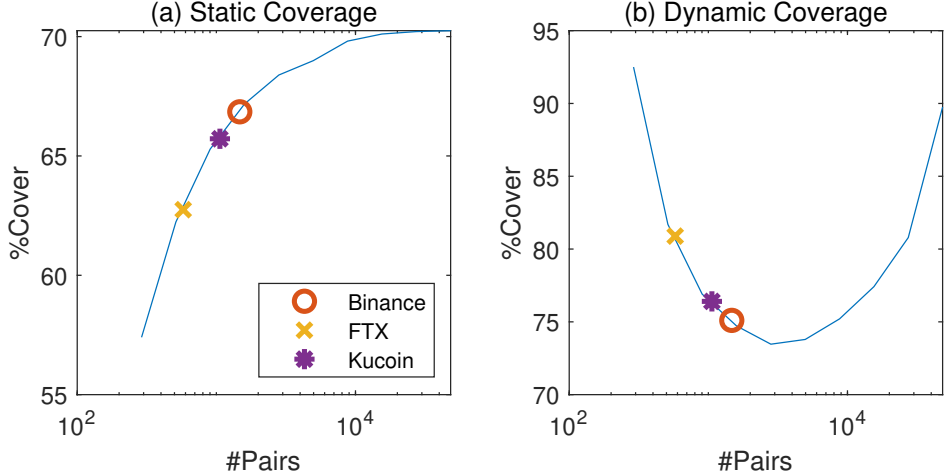}
	\caption{a) Under different $M$ limits, the percentage of traffic covered. b) Under different $M$ limits, for the optimal trading pair in the previous cycle, the average retention ratio in the next cycle. The curve is drawn on the basis of Binance, and the horizontal positions of FTX and Kucoin are given by $N_B/N_{F or K}*M_{F or K}$.}
	
\end{figure}\label{fig:cover}

Next, we examine how much $M$ is appropriate. It is true that the larger the $M$ is, the more transactions can be directly carried, as shown in Fig. \ref{fig:cover}(a). However, the smaller the trading pair that is carried, the more volatile its flow is. That is to say, in the case where the settable trading pairs can always only account for a small part of all possible trading pairs, that is, $M\ll \frac{N(N-1)}{2}$, then the current optimal solution $\mathbf{G}^*$ is destined to face frequent adjustments to keep up with the rhythm of the market's rotation of hot topics. As shown in Fig. \ref{fig:cover}(b), the stability gradually increases only after $M$ reaches a certain value; however, this interval is usually difficult to reach. Comparing the three exchanges, we found that Binance has more trading pairs, which means more frequent adjustments; while FTX and Kucoin are much more stable. In practice, exchanges need to balance coverage and adjustment frequency to ensure users have a stable trading experience.

	\section{Conclusion}
	
	This article examines how to automatically set up optimal trading pairs on crypto exchanges. It provides some quantitative basis for evaluating the problem of trading pair setup, and proposes to search for an optimal solution by using a variant of SVD in conjunction with the branch and bound method. After simulating on historical data, we found that the total number of trading pairs should be maintained at a reasonable ratio.
	
	If we assume that the orderbook-based market mechanism will remain in the future, there are different needs for this process for humans and machines. For humans, it may be desirable to control the number of quotes so that there is no need to query for the existence of a trading pair beforehand. But for machines, restrictions on quotes can hurt transaction efficiency. In the future, we envision a two-tier trading market structure, where human traders only need to propose the source, target and quantity of the conversion, and the program will automatically find the optimal exchange path. Likewise, for the trading platform, the setting of the optimal trading pair is done automatically. It is foreseeable that this will be a more efficient transaction mechanism than the current manual method.
	
\bibliographystyle{plain}
\bibliography{paper}  

\begin{thebibliography}{10}

\bibitem{byrd2000trust}
Richard~H Byrd, Jean~Charles Gilbert, and Jorge Nocedal.
\newblock A trust region method based on interior point techniques for
  nonlinear programming.
\newblock {\em Mathematical programming}, 89(1):149--185, 2000.

\bibitem{Cohen2021ATO}
Michael~X. Cohen.
\newblock A tutorial on generalized eigendecomposition for source separation in
  multichannel electrophysiology.
\newblock 2021.

\bibitem{egu2020comparable}
Oscar Egu and Patrick Bonnel.
\newblock How comparable are origin-destination matrices estimated from
  automatic fare collection, origin-destination surveys and household travel
  survey? an empirical investigation in lyon.
\newblock {\em Transportation Research Part A: Policy and Practice},
  138:267--282, 2020.

\bibitem{fang2021ascertaining}
Fan Fang, Waichung Chung, Carmine Ventre, Michail Basios, Leslie Kanthan,
  Lingbo Li, and Fan Wu.
\newblock Ascertaining price formation in cryptocurrency markets with machine
  learning.
\newblock {\em The European Journal of Finance}, pages 1--23, 2021.

\bibitem{fang2022cryptocurrency}
Fan Fang, Carmine Ventre, Michail Basios, Leslie Kanthan, David Martinez-Rego,
  Fan Wu, and Lingbo Li.
\newblock Cryptocurrency trading: a comprehensive survey.
\newblock {\em Financial Innovation}, 8(1):1--59, 2022.

\bibitem{fekih2021data}
Mariem Fekih, Tom Bellemans, Zbigniew Smoreda, Patrick Bonnel, Angelo Furno,
  and St{\'e}phane Galland.
\newblock A data-driven approach for origin--destination matrix construction
  from cellular network signalling data: a case study of lyon region (france).
\newblock {\em Transportation}, 48:1671--1702, 2021.

\bibitem{fil2020pairs}
Miroslav Fil and Ladislav Kristoufek.
\newblock Pairs trading in cryptocurrency markets.
\newblock {\em IEEE Access}, 8:172644--172651, 2020.

\bibitem{krishnakumari2020data}
Panchamy Krishnakumari, Hans Van~Lint, Tamara Djukic, and Oded Cats.
\newblock A data driven method for od matrix estimation.
\newblock {\em Transportation Research Part C: Emerging Technologies},
  113:38--56, 2020.

\bibitem{Nakamoto2008BitcoinAP}
Satoshi Nakamoto.
\newblock Bitcoin: A peer-to-peer electronic cash system.
\newblock 2008.

\bibitem{pastor2005solving}
Franck Pastor and Etienne Loute.
\newblock Solving limit analysis problems: an interior-point method.
\newblock {\em Communications in numerical methods in engineering},
  21(11):631--642, 2005.

\bibitem{ramos2017introducing}
Jos{\'e}~Pedro Ramos-Requena, JE~Trinidad-Segovia, and MA~S{\'a}nchez-Granero.
\newblock Introducing hurst exponent in pair trading.
\newblock {\em Physica A: statistical mechanics and its applications},
  488:39--45, 2017.

\bibitem{ramos2020some}
Jos{\'e}~Pedro Ramos-Requena, Juan~Evangelista Trinidad-Segovia, and
  Miguel~{\'A}ngel S{\'a}nchez-Granero.
\newblock Some notes on the formation of a pair in pairs trading.
\newblock {\em Mathematics}, 8(3):348, 2020.

\bibitem{ros2022practical}
Xavier Ros-Roca, L{\'\i}dia Montero, Jaume Barcel{\'o}, Klaus N{\"o}kel, and
  Guido Gentile.
\newblock A practical approach to assignment-free dynamic origin--destination
  matrix estimation problem.
\newblock {\em Transportation Research Part C: Emerging Technologies},
  134:103477, 2022.

\bibitem{tomazella2020comprehensive}
Caio~Paziani Tomazella and Marcelo~Seido Nagano.
\newblock A comprehensive review of branch-and-bound algorithms: Guidelines and
  directions for further research on the flowshop scheduling problem.
\newblock {\em Expert Systems with Applications}, 158:113556, 2020.

\bibitem{Wood2014ETHEREUMAS}
Daniel~Davis Wood.
\newblock Ethereum: A secure decentralised generalised transaction ledger.
\newblock 2014.

\bibitem{Wang2020ResearchOR}
Zhi yuan Wang, Haiting Chen, and Yang mei Ou.
\newblock Research on regional traffic and economic linkage based on
  accessibility and gravity model--taking hengyang, china as an example.
\newblock {\em IOP Conference Series: Earth and Environmental Science}, 510,
  2020.

\end{thebibliography}
	
\end{document}